\documentclass[prl,twocolumn,superscriptaddress]{revtex4-2}
\usepackage{amsfonts}
\usepackage{amssymb}
\usepackage{amsmath}
\usepackage{graphicx}
\usepackage{dcolumn}
\usepackage{bm}
\usepackage{units}
\usepackage{multirow}
\usepackage{CJKutf8}
\usepackage{subfigure}
\usepackage{color}%
\usepackage{soul}
\usepackage{ulem}
\usepackage{url}
\usepackage[colorlinks,linkcolor=blue,anchorcolor=blue,citecolor=blue,urlcolor=blue]{hyperref}
\usepackage{array}
\begin{document}

\title{Skyrmion size in skyrmion crystals} 
\author{H. T. Wu}
\affiliation{Physics Department, The Hong Kong University of Science 
and Technology (HKUST), Clear Water Bay, Kowloon, Hong Kong}
\affiliation{HKUST Shenzhen Research Institute, Shenzhen 518057, China}
\author{X. C. Hu}
\affiliation{Physics Department, The Hong Kong University of Science 
and Technology (HKUST), Clear Water Bay, Kowloon, Hong Kong}
\affiliation{HKUST Shenzhen Research Institute, Shenzhen 518057, China}
\author{K. Y. Jing}
\affiliation{Physics Department, The Hong Kong University of Science 
and Technology (HKUST), Clear Water Bay, Kowloon, Hong Kong}
\affiliation{HKUST Shenzhen Research Institute, Shenzhen 518057, China}
\author{X. R. Wang}
\email[Corresponding author: ]{phxwan@ust.hk}
\affiliation{Physics Department, The Hong Kong University of Science 
and Technology (HKUST), Clear Water Bay, Kowloon, Hong Kong}
\affiliation{HKUST Shenzhen Research Institute, Shenzhen 518057, China}
\date{\today}

\begin{abstract}
A magnetic skyrmion is a topological object that can exist as a solitary embedded in 
the vast ferromagnetic phase, or coexists with a group of its ``siblings'' in various 
stripy phases as well as skyrmion crystals (SkXs). Isolated skyrmions and skyrmions in 
an SkX are circular while a skyrmion in other phases is a stripe of various forms. 
Unexpectedly, the sizes of the three different types of skyrmions depend on material  
parameters differently. For chiral magnetic films with exchange stiffness constant $A$, 
the Dzyaloshinskii-Moriya interaction (DMI) strength $D$, and perpendicular magnetic 
anisotropy $K$, $\kappa\equiv\pi^2D^2/(16AK)=1$ separates isolated skyrmions from 
condensed skyrmion states. In contrast to isolated skyrmions whose size increases with 
$D/K$ and is insensitive to $\kappa\ll1$ and stripe skyrmions whose width increases 
with $A/D$ and is insensitive to $\kappa\gg1$, the size of skyrmions in SkXs is inversely 
proportional to the square root of skyrmion number density and decreases with $A/D$. 
This finding has important implications in our search for stable smaller skyrmions at 
the room temperature in applications. 
\end{abstract}

\maketitle
Skyrmions, localized topological objects characterized by the skyrmion number $Q=\frac{1}{4\pi}
\int{\mathbf m}\cdot(\partial_x {\mathbf m}\times \partial_y {\mathbf m}){\rm\,d}x{\rm d}y$, 
were originally proposed as resonance states of baryons \cite{skyrme}, but were first 
unambiguously observed in chiral magnets that involve the Dzyaloshinskii-Moriya interaction 
(DMI) \cite{Robler,Muhlbauer,Yu,Onose,Park,Heinze,Romming,Li,Jiang,news,Woo,Tian,Yuan}.
Here $\mathbf m$ is the unit vector of magnetization and $Q$ measures the number of 
times that spins, after moving to the origin, cover the unit sphere of $|\mathbf m|=1$. 
Skyrmions have received intensive and extensive studies \cite{Robler,Muhlbauer,Yu,Onose,
Park,Heinze,Romming,Li,Jiang,news,Woo,roadmap,xrw3,xrw4,Jiang,BandK,Romming,size2015,Karhu,
PdFeIr,Tian,Yuan,Nagaosa,Fert,ZhouYan,size2016,Xu,Li,Yuan2018,Iwasaki,Fert,
Xiansi,Wiesendanger,temp_grad,Ukleev,MnSi_anis,Romming,size2015,gong_prb_2020,Thiaville2013,
gong_prb_2020,Yuan2018} in magnetics for their academic interest and promising 
applications in information technology.  

Our knowledge about magnetic skyrmions has been greatly advanced in recent years 
\cite{roadmap,xrw3,xrw4}. Skyrmions in metastable states are fundamentally different 
from the ones in the ground state. A skyrmion is circular and its formation energy is 
positive when the skyrmion is metastable and ferromagnetic ordering is the ground state. 
Skyrmion morphologies are various stripes that could be ramified or non-ramified, 
long or short, and straight or curved when skyrmions are the ground state. 
Thus, present knowledge of skyrmion zoo consisting of three types of circular skyrmions 
\cite{roadmap}, namely Bloch skyrmions (whirl skyrmions), the N\'{e}el skyrmions 
(hedgehog skyrmions), and anti-skyrmions, should be greatly expanded by including 
various stripes in disordered stripy phases, in helical states, and in maze. 
Even when skyrmions are the ground state, individual skyrmion can still be disk-like 
in triangular lattice of closely packed skyrmions where skyrmion-skyrmion repulsion 
dominates the morphology of individual skyrmions. It is already known \cite{Xiansi} 
that the size of isolated skyrmions increases with the DMI strength, and decreases with 
the exchange stiffness coefficient and magnetic anisotropy \cite{Xiansi,Wiesendanger}. 
It is also known \cite{xrw3} that the width of stripe skyrmions increases with exchange 
stiffness coefficient and decreases with the DMI strength, but does not depend on 
skyrmion number density. Since the shape of a skyrmion in a skyrmion crystal (SkX) has 
different underlying physics as an isolated skyrmion and a stripe skyrmion, the 
parameter dependence of the size of disk-like skyrmions in an SkX may also be different 
although people do not distinguish a skyrmion in an SkX from an isolated ones to date. 

In this paper, we use a generic chiral magnetic film with exchange stiffness 
constant $A$, the DMI strength $D$, and perpendicular magnetic anisotropy $K$ to 
show that $\kappa\equiv \pi^2D^2/(16AK)<1$ and $\kappa>1$ determines the existence 
of isolated skyrmions (the former) and stripe skyrmions or SkXs (the latter).  
Although skyrmions in SkXs and stripe skyrmions are ``siblings", the dominate 
repulsion among skyrmions in SkX changes the parameter dependence of skyrmion size.
This finding provides a general guidance in skyrmion size manipulation.
	
We consider a two-dimensional (2D) ferromagnetic film of thickness $d$ in the 
$xy$-plane described by the following energy
\begin{equation}
\begin{aligned}
E=&d\iint \lbrace A|\nabla \mathbf{m}|^2+K(1-m_z^2)
\\&+D[(\mathbf{m}\cdot\nabla)m_z-m_z\nabla\cdot\mathbf{m}]\rbrace \mathrm{\,d}S,
\end{aligned}
\label{energy}
\end{equation}
where $\mu_0$ is the vacuum permeability. The energy is set to zero, $E=0$, for 
ferromagnetic state of $|m_z|=1$. Films with $\kappa<1$ support isolated circular 
skyrmions that are metastable state of energy $8\pi Ad\sqrt{1-\kappa}$ \cite{Xiansi}. 
The dynamics of $\mathbf{m}$ is governed by the Landau-Lifshitz-Gilbert (LLG) equation,
\begin{equation}
\frac{\partial \mathbf{m}}{\partial t} =-\gamma\mathbf{m} \times \mathbf{H}_{\rm eff} +
\alpha \mathbf{m} \times \frac{\partial \mathbf{m}}{\partial t}, 
\label{llg}
\end{equation}
where $\gamma$ and $\alpha$ are respectively gyromagnetic ratio and the Gilbert damping constant. 
The effective magnetic field $\mathbf{H}_{\rm eff}=2A\nabla^2\mathbf{m}+2K_1m_z\hat z+
\mathbf{H}_{\rm\,d}+\mathbf{H}_{\rm DM}$ includes the exchange field, the magneto-crystalline 
anisotropy field of coefficient $K_1$, the external magnetic field, the demagnetizing field 
$\mathbf{H}_{\rm\,d}$, and the DMI field $\mathbf{H}_{\rm DM}$ respectively. 
$\mathbf{H}_{\rm\,d}$ can be included in the effective anisotropy of $K=K_1-\mu_0M_{\rm s}^2/2$. 
This approximation is good when $d$ is much smaller than the exchange length \cite{Xiansi}.

In the absence of energy source such as an electric current and at zero temperature, the LLG 
equation describes a dissipative system whose energy can only decrease \cite{xrw1,xrw2}. 
Thus, the long-time solution of the LLG equation with any initial configuration must be a stable 
spin texture of Hamiltonian Eq.~\eqref{energy}. In this study, we choose three groups of model 
parameters to simulate those chiral magnets such as MnSi with $A=0.27\sim 0.4\,\hbox{pJ/m}$, $D=0.15
\sim 0.33\,\hbox{mJ/m}^2$, $K=0.02 \sim 0.2\,\hbox{MJ/m}^3$, $M_{\rm s}=0.15\sim 0.5\,\hbox{MA/m}$ 
\cite{Karhu}; $\hbox{Co-Zn-Mn}$ with $A=5\sim 11\,\hbox{pJ/m}$, $D=0.5 \sim 2\,\hbox{mJ/m}^2$, 
$K=20\sim 80\,\hbox{kJ/m}^3$, $M_{\rm s}=0.15\sim 0.35\,\hbox{MA/m}$ \cite{Ukleev}; and PdFe/Ir 
and W/Co$_{20} $Fe$_{60}$B$_{20}$/MgO \cite{Jiang,BandK,Romming,size2015,Fert,PdFeIr} with $A=2\sim 10
\,\hbox{pJ/m}$, $D=0.68 \sim 4\,\hbox{mJ/m}^2$, $K=0.00228 \sim 5\,\hbox{MJ/m}^3$, $M_{\rm s}=650\sim 1160\,\hbox{kA/m}$.
Periodic boundary conditions are used to eliminate boundary effects and the MuMax3 package \cite{MuMax3} 
is employed to numerically solve the LLG equation with mesh size of 
\rm{$1\,{\rm nm}\times1\,{\rm nm}\times1\,{\rm nm}$}.
The number of stable states and their structures should not depend on the Gilbert damping constant. 
We use a large $\alpha=0.25$ to speed up our simulations. 

\begin{figure}
\centering
\includegraphics[width=8.5cm]{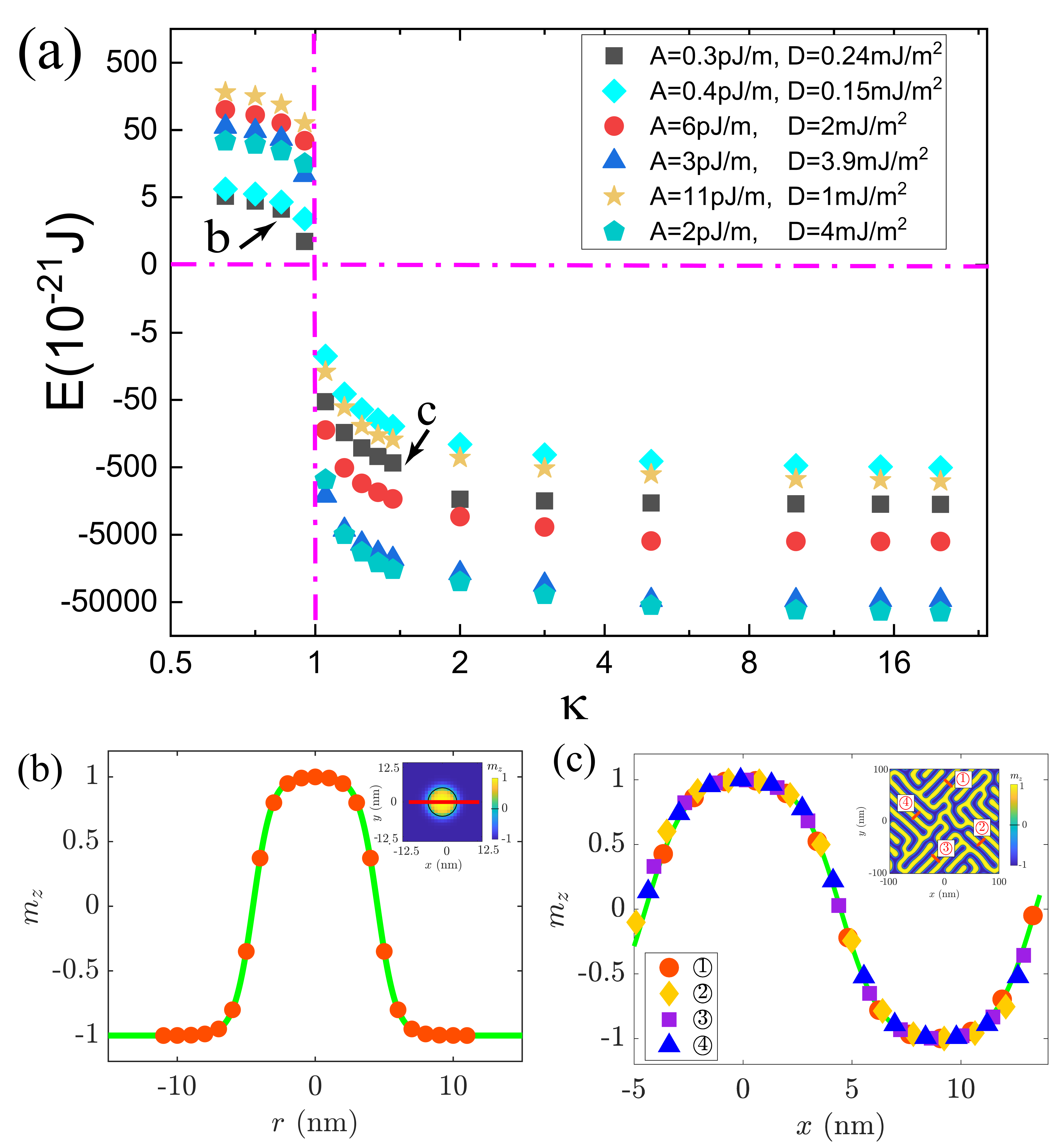}\\
\caption{a) Energy (in ${\rm arcsinh}(E)$ scale) as a function of $\kappa$ (in $\log(\kappa)$ scale) 
for various model parameters with very different $A$, $D$ and $K$. $E$ changes sign at $\kappa=1$ 
that separate isolated skyrmions from stripe skyrmions and skyrmions in SkXs. 
(b) Spin profile $m_z(r)$ of an isolated skyrmion shown in the inset with parameters denoted by $b$ in (a). 
Symbols are simulation data and curve are theoretical fit with $R=4.513\,$nm and $w=1.356\,$nm. 
(c) Spin profile $m_z(x)$ of stripe skyrmion with parameters denoted by $c$ in (a). 
Symbols are simulation data for stripes \textcircled{1}-\textcircled{4} in the inset and the solid 
curve is the theoretical fit with $L=8.95\,$nm and $w=1.82\,$nm. $x$ is measured across the stripe 
and $x=0$ is the stripe centre.}
\label{fig1}
\end{figure}
\par

Our previous studies \cite{xrw3,xrw4} demonstrate that a small nucleation domain of $m_z=-1$ in the 
background of ferromagnetic phase of $m_z=1$ can develop into a skyrmion in a chiral magnetic film. 
Using MuMax3 \cite{MuMax3}, we obtain the energy of one static skyrmion. The sign of $E$ tells us 
whether the skyrmion is a metastable state or a ground state because $E=0$ is set for the ferromagnetic 
state of $|m_z|=1$. Figure 1(a) plots the energy of one skyrmion in chiral magnets as a function of $\kappa$. 
The squares and diamonds, circles and stars, up-triangles and pentagons, are respectively from three 
groups of model parameters listed above. Because the huge range of $E$ is covered around zero, $E$ 
is plotted in ${\rm arcsinh}(E)$ scale and $\kappa$ is in $\log(\kappa)$ scale. 
Clearly, $\kappa=1$ separate metastable circular isolated skyrmions from stripe skyrmions.
 
The spin profile of isolated skyrmions can be approximated by $\Theta (r)=2\arctan\left[\frac{\sinh(r/w)}
{\sinh(R/w)}\right]$ \cite{Xiansi}. $\Theta$ is the polar angle of the magnetization at position $r$ measured 
from the skyrmion centre. $R$ and $w$ measure respectively the skyrmion size and skyrmion wall thickness. 
Figure \ref{fig1}(b) shows the comparison of spin profile from MuMax3 (symbols) and theoretical fit with 
$R=4.513\,$nm and $w=1.356\,$nm for an isolated skyrmion with $A=0.4\,\hbox{pJ/m}$, $D=0.15\,\hbox{mJ/m}^2$, 
$K=0.2 \,\hbox{MJ/m}^3$, $M_{\rm s}=0.15\,\hbox{MA/m}$. The inset shows the skyrmion structure. 
This profile leads to the skyrmion size formula of $R=w/\sqrt {1-\kappa}$ and $w=\pi D/(4K)$ \cite{Xiansi}. 

\begin{figure*}
\centering
\includegraphics[width=17cm]{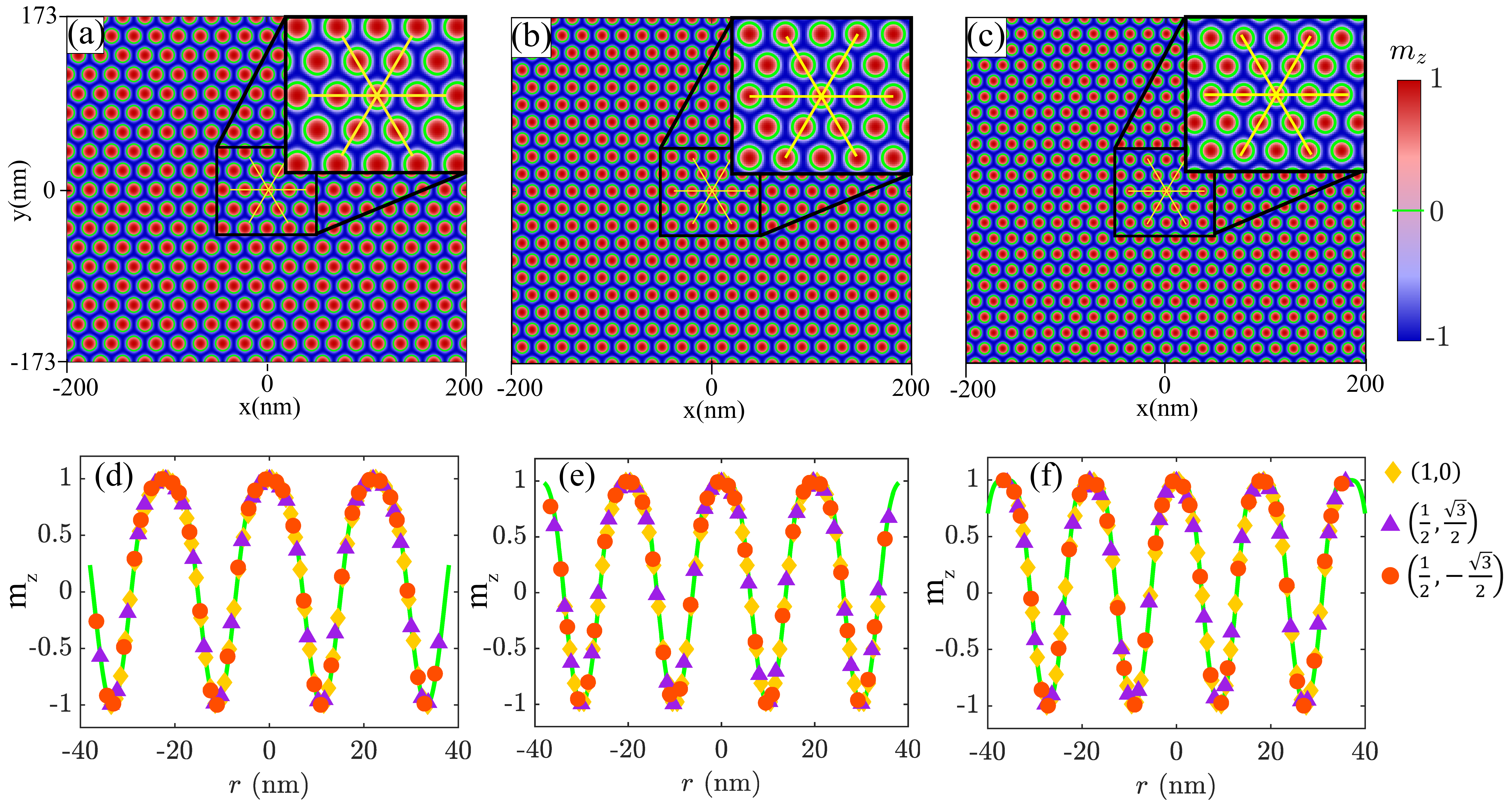}\\
\caption{SkXs and skyrmion spin profiles at various skyrmion number density 
$2.34\times10^{-3}\hbox{/nm}^2$ (a, d), $2.89\times10^{-3}\hbox{/nm}^2$ (b, e), and 
$3.50\times10^{-3}\hbox{/nm}^2$ (c, f). (d-f) Spin profiles $m_z(r)$ along the red 
lines in (a), (b), and (c), respectively. The symbols are numerical data and the solid 
lines are the theoretical fits with $R=7.042\,$nm (d), $6.144\,$nm (e), and $5.399\,$nm (f).}
\label{fig2}
\end{figure*}

The spin profile of stripe skyrmions can be approximated by $\Theta (x)=2\arctan\left[\frac{\sinh(L/2w)}
{\sinh(|x|/w)}\right]$ for $m_z<0$ and $\Theta (x)=2\arctan\left[\frac{\sinh(|x|/w)}{\sinh(L/2w)}\right]$ 
for $m_z>0$ \cite{xrw3}. $x$ is measured across the stripe and $x=0$ is at stripe centre. 
$L$ and $w$ measure respectively the stripe width and skyrmion wall thickness. Figure \ref{fig1}(c) is spin 
profile of stripe skyrmion denoted by $c$ in Fig. 1(a) with model parameter of $A=0.3\,\hbox{pJ/m}$, 
$D=0.24 \,\hbox{mJ/m}^2$, $K=0.096 \,\hbox{MJ/m}^3$, $M_{\rm s}=0.15 \,\hbox{MA/m}$. 
The symbols are the numerical data from various locations of stripe skyrmion 
shown in the inset and the solid curve is the theoretical fit with $L=8.95$ nm and $w=1.82$ nm. 
The excellent agreements between the numerical data and the approximate profiles demonstrate 
that one can use this theoretical profile to extract the skyrmion size accurately. 
All data from different stripes fall on the same curve demonstrate that stripes, building blocks of 
irregular skyrmions, are identical. The excellent stripe skyrmion profile leads to the stripe width 
formula $L=g(\kappa) A/D$, where $g(\kappa)$ is about 6.23 for $\kappa\gg 1$ \cite{xrw3}, 
opposite dependence on $D$ as that for an isolated skyrmion. 

\begin{figure}
\centering
\includegraphics[width=8.5cm]{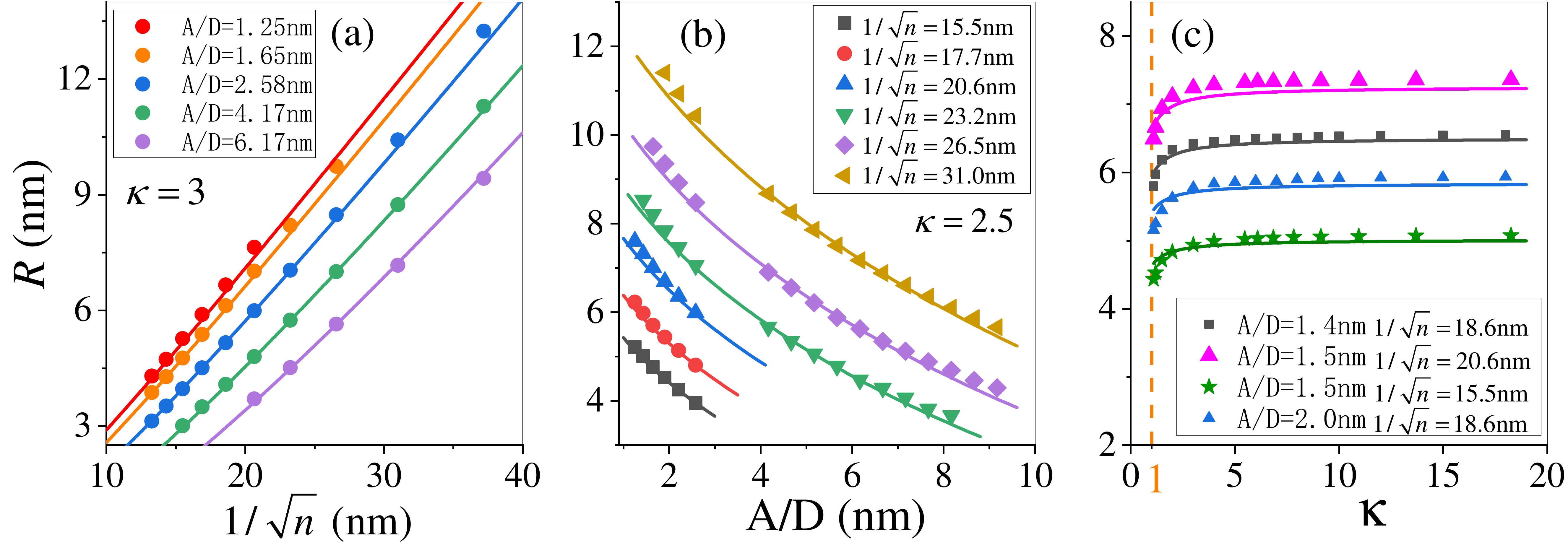}\\
\caption{Skyrmion size in SkXs as a function of skyrmion number density $n$ for various $A/D$ 
and fixed $\kappa$ (a), as a function of $A/D$ for various $n$ and fixed $\kappa$ (b), and as 
a function of $\kappa$ for various $n$ and $A/D$. The symbols are numerical data from MuMax3. 
The solid lines are the formula of Eq.~\eqref{size}.}
\label{fig3}
\end{figure}

In order to study skyrmions in SkXs, we consider model parameters satisfying $\kappa>1$ below. 
All initial configurations have sufficient number of nucleation domains in triangular lattices. 
Each domain of $m_z=-1$ is a disk of 10nm in diameter. Each initial configuration leads to an 
SkX as shown in Figs. \ref{fig2}(a)-(c) with respectively 324, 400, and 484 skyrmions in a 
film of $400\,\hbox{nm}\times 346\,\hbox{nm}\times 1\,\hbox{nm}$. The model parameters are $A=5 
\,\hbox{pJ/m}$, $D=3\,\hbox{mJ/m}^2$, $K=4.87\,\hbox{kJ/m}^3$, $M_{\rm s}=0.2\,\hbox{MA/m}$.  
Figs. \ref{fig2}(a)-(c) show that the territories of $m_z>0$ and $m_z<0$ are different, and have 
different spin distribution. Lattice constant and skyrmion size are two obvious length scales. 
Spin profiles of isolated skyrmions and stripe skyrmions should not work for SkXs because of 
different underlying physics, a new spin polar angle distribution of $\Theta (r)=2\arctan\left\{
\frac{\tan\left[\frac{\pi}{2}\cos(\frac{\pi R}{L})\right]}{\tan\left[\frac{\pi}{2}\cos(\frac{\pi 
r}{L})\right]}\right\}$ along the three triangular lattice directions, {(1,0), (1/2, $\sqrt{3}/2$), 
and (1/2, $-\sqrt{3}/2$), is proposed, where $r$ labels the points on the yellow lines in (a), 
(b), and (c) with $r=0$ being the centre of one chosen skyrmion. 
$R$ and $L$ are respectively the skyrmion size (the radius of $m_z=0$ contour) and lattice constant. 
$L$ relates to skyrmion number density $n$ in a triangular SkX as $L\approx 1.07/\sqrt{n}= 22.11\,$nm 
(a), $19.90\,$nm (b), and $18.09\,$nm (c). Figure \ref{fig2}(d), (e), and (f) are $m_z(r)$ along the yellow lines 
in (a), (b), and (c), respectively. Symbols are numerical data from MuMax3 while the curves are the 
fitting of $m_z(r)=\cos\Theta$ with the fitting parameter $R=7.042\,$nm (d), $6.144\,$nm (e), $5.399\,$nm (f).
One of the interesting observations is that $R$ depends only on $n$, $A/D$, and $\kappa$. 
Numerical evidences on the result can be found in the Supplementary Materials.

If we assume energy per skyrmion in an SkX is from a skyrmion whose spin texture is described by 
$\Theta (r)$ for $0\le r \le L/2$, then energy per skyrmion can be expressed as a function of 
variable $R$,  
\begin{equation}
E(R)=2\pi d[Af_1(\varepsilon)+DLf_2(\varepsilon)+KL^2f_3(\varepsilon)],
\label{energy1}
\end{equation}
where $\varepsilon=\frac{2R}{L}$, $f_1(\varepsilon)=4f^2(\varepsilon){\displaystyle\int_{0}^{1}}\frac{\left[\xi^2f'^2(\varepsilon)+
f^2(\xi)\right]}{\xi\left[f^2(\xi)+f^2(\varepsilon)\right]^2}{\rm\,d}\xi$, 
$f_2(\varepsilon)=f(\varepsilon){\displaystyle\int_{0}^{1}}\left\{\frac{\xi f'(\xi)}{f^2(\xi)+f^2(\varepsilon)}+
\frac{f(\xi)\left[f^2(\varepsilon)-f^2(\xi)\right]}{\left[f^2(\xi)+f^2(\varepsilon)\right]^2}\right\}
{\rm\,d}\xi$, $f_3(\varepsilon) ={\displaystyle\int_{0}^{1}}\left[\frac{f(\xi)f(\epsilon)}{f^2(\xi)+f^2(\varepsilon)}
\right]^2{\rm\,d}\xi$, and $f(x)=\tan\left[\frac{\pi}{2}\cos(\frac{\pi }{2}x)\right]$. 
The skyrmion size is then the solution of $\frac{{\rm\,d}E}{{\rm\,d}R}=0$.
Under certain reasonable assumptions (see the Supplementary Materials), one has  
\begin{equation}
R(n,A/D,\kappa)=\frac{0.54}{\sqrt{n}}-\sqrt{\frac{0.51}{\sqrt{n}\frac{A}{D}+0.56n\frac{A^2}{D^2}-0.02\frac{1}{\kappa}}}\frac{A}{D}.
\label{size}
\end{equation}
Unlike the width of stripe skyrmions that does not depend on $n$ and increase with $A/D$, 
Eq.~\eqref{size} says that $R$ is almost inversely proportional to $\sqrt{n}$ and decreases with $A/D$. 
The interesting dependences agree with the underlying physics of SkX that is the balance of 
skyrmion-skyrmion repulsion and intrinsic stripe nature of skyrmions. 
$R$ should then be proportional to the lattice constant of a SkX that is the skyrmion-skyrmion 
distance and, in turn, is inversely proportional to $\sqrt{n}$. Since skyrmions in a SkX come from 
the squeeze of stripe skyrmions, the amount of width deduction from the squeeze should be proportional 
to the original stripe width proportional to $A/D$. Thus, $R$ should not only be smaller than the 
natural stripe width, but also decreases with $A/D$ as what Eq.~\eqref{size} says. Of course, one 
should not expect coefficients of $1/\sqrt{n}$ and $A/D$ to be very accurate because the obvious 
approximations involved that include impossible filling by disks of radius $L/2$ and the neglect of 
energy from uncovered region in Eq. \eqref{energy1}.

We carried out MuMax3 simulations with very different $n$, $A$, $D$, $K$ in three different parameter 
regions. For each fixed parameters, skyrmion size $R$ in the corresponding SkX can be obtained either 
from the size of $m_z=0$ contour or through fitting of numerical spin profile to $\Theta (r)$. 
The difference between two extracted values is negligible (see the Supplementary Materials), and all 
data shown below is from contour measurement. One can test our theoretical predictions of 
Eq.~\eqref{size} against numerical simulations. As mentioned above, simulations show that $R$ depends 
only on $n$, $A/D$, and $\kappa$. Fig. \ref{fig3}(a) plots $R$ vs $1/\sqrt{n}$ for various fixed $A/D$ 
and $\kappa$ from three groups model parameters. Simulation results (symbols) are well described by 
our formula (solid curves) of Eq.~\eqref{size} without any fitting. $R$ is almost linear in $1/\sqrt{n}$}. 
The $A/D$ dependence and $\kappa$ dependence of $R$ can also be captured well by Eq.~\eqref{size} as 
shown by good agreement between simulation data (symbols) and theoretical formula of Eq.~\eqref{size} 
(the solid curve) in Fig. \ref{fig3}(b) ($A/D$) and Fig.~\ref{fig3}(c) ($\kappa$), respectively.
Interestingly, Eq. \eqref{size} says that $R$ does not depend on anisotropy $K$ for $\kappa\gg 1$.
Our new spin profile captures successfully the skyrmion-skyrmion repulsion effect that requires 
$R$ to be proportional to the lattice constant, the inverse of the skyrmion number density $n$. 
%The nice agreement between our theory, Eq. \eqref{size}, and the numerical results from MuMax3 
%demonstrates that $E(R)$ is mainly from  energy from the unfilled area of three closely packed disks of 
%radius $L/2$ is negligible to .

Magnetic skyrmions can be isolated, or condensed regular or irregular stripes distributed 
randomly or arranged as helical states of one-dimensional stripe lattices, or condensed circular 
skyrmions in crystal forms. The irregular stripes can even be ramified in dendrite or maze forms.  
The underlying physics of skyrmion size are different for isolated skyrmions, 
skyrmions in an SkX and stripe skyrmions in vastly different morphologies and structures. 
It is known that skyrmion formation energy are positive for isolated skyrmions and negative for 
stripe skyrmions and SkXs. Positive formation energy results in an extra energy for skyrmion 
surfaces, similar to the surface tension of a liquid droplet, that favours a circular shape. 
Negative skyrmion formation energy is similar to negative surface tension of a liquid droplet 
that prefers to wet its contacts and spread out. From this viewpoint, it may not be surprising 
to see different parameter dependences of skyrmion size for isolated skyrmions and skyrmions 
in SkXs or stripe skyrmions. It is also not surprising that spin profile of skyrmions in SkXs 
has little similarity as those of isolated and stripe skyrmions because their shape comes 
from the balance of the skyrmion squeezing and their stripe nature. Two nearby skyrmions 
start to repel each other when their distance is order of $2L\approx 13 A/D$. 
Thus, the critical skyrmion number density for SkX is about $n_c\approx D^2/(169A^2)$. 
For $n>n_c$, skyrmions tend to form closely packed triangular lattice with lattice constant of 
$1.07/\sqrt{n}$. Due to the squeezing of skyrmions, the skyrmion size (diameter of $m_z=0$ 
contour) should be smaller than $0.53/\sqrt{n}$, this squeezing effect leads to Eq. \ref{size}  
in which skyrmion size decreases with $A/D$, opposite to the parameter dependence of the width 
of stripe skyrmions.

In applications, creating stable skyrmions of smaller size at the room temperature is a goal.
According to current results, one needs to distinguish an isolated skyrmion ($\kappa<1$) 
from a skyrmion in condensed skyrmion phase ($\kappa>1$). In the case of $\kappa<1$, materials 
with large $A$ and smaller $D$ are preferred since larger $A$ and smaller $D$ imply a higher 
$T_{\rm c}$ and smaller skyrmion size. On the other hand, in the case of $\kappa>1$, materials 
with larger $A$ and larger $D$ are required so that one may have both higher $T_{\rm c}$ and 
smaller skyrmion size.

In conclusion, we clarify that there are three distinct skyrmions, namely circular 
isolated skyrmions, stripe skyrmions, and disk-like skyrmions in SkXs. The material 
parameter dependences of skyrmion size are very different. Unlike an isolated skyrmion 
whose size increases with $D/K$ and $\kappa<1$, or a stripe skyrmion whose size increases 
with $A/D$ and decreases with $\kappa>1$, the size of a disk-like skyrmion in SkXs is 
inversely proportional to the skyrmion number density and decreases with $A/D$. Skyrmion 
size is an very important quantity, and high-density data storage require smaller 
skyrmions in applications. The findings here provide a practical guidance to search 
proper materials in different applications. 

\begin{acknowledgements}
This work is supported by National Key R\&D Program of China, grant No 2020YFA0309600, 
the NSFC Grant (No. 11974296 and 11774296) and Hong Kong RGC Grants (No. 16301518 and 
16301619). Partial support by the National Key Research and Development Program of China 
(Grant No. 2018YFB0407600) is also acknowledged. 
\end{acknowledgements}

\end{document}